%
% the following is to use blackboard bold fonts --
\let\useblackboard=\iftrue
%
% activate this if you don't have them.
%\let\useblackboard=\iffalse
%
% You might also need to remove this line.
\newfam\black
\input harvmac.tex
\def\Title#1#2{\rightline{#1}
\ifx\answ\bigans\nopagenumbers\pageno0\vskip1in%
\baselineskip 15pt plus 1pt minus 1pt
\else%\special{papersize=11in,8.5in}%
\def\listrefs{\footatend\vskip 1in\immediate\closeout\rfile\writestoppt
\baselineskip=14pt\centerline{{\bf References}}\bigskip{\frenchspacing%
\parindent=20pt\escapechar=` \input
refs.tmp\vfill\eject}\nonfrenchspacing}
\pageno1\vskip.8in\fi \centerline{\titlefont #2}\vskip .5in}

\ifx\answ\bigans\def\tcbreak#1{}\else\def\tcbreak#1{\cr&{#1}}\fi
\useblackboard
\message{If you do not have msbm (blackboard bold) fonts,}
\message{change the option at the top of the tex file.}
\font\blackboard=msbm10 scaled \magstep1
\font\blackboards=msbm7
\font\blackboardss=msbm5
%\newfam\black
\textfont\black=\blackboard
\scriptfont\black=\blackboards
\scriptscriptfont\black=\blackboardss

\else

\fi
% *************************************
%
\def\yboxit#1#2{\vbox{\hrule height #1 \hbox{\vrule width #1
\vbox{#2}\vrule width #1 }\hrule height #1 }}
\def\fillbox#1{\hbox to #1{\vbox to #1{\vfil}\hfil}}
\def\ybox{{\lower 1.3pt \yboxit{0.4pt}{\fillbox{8pt}}\hskip-0.2pt}}

\def\comments#1{}

\def\II{\relax{I\kern-.07em I}}

\def\IZ{\relax\ifmmode\mathchoice
{\hbox{\cmss Z\kern-.4em Z}}{\hbox{\cmss Z\kern-.4em Z}}
{\lower.9pt\hbox{\cmsss Z\kern-.4em Z}}
{\lower1.2pt\hbox{\cmsss Z\kern-.4em Z}}\else{\cmss Z\kern-.4em
Z}\fi}
\def\IB{\relax{\rm I\kern-.18em B}}
\def\IC{{\relax\hbox{$\inbar\kern-.3em{\rm C}$}}}
\def\ID{\relax{\rm I\kern-.18em D}}
\def\IE{\relax{\rm I\kern-.18em E}}
\def\IF{\relax{\rm I\kern-.18em F}}
\def\IG{\relax\hbox{$\inbar\kern-.3em{\rm G}$}}
\def\IGa{\relax\hbox{${\rm I}\kern-.18em\Gamma$}}
\def\IH{\relax{\rm I\kern-.18em H}}
\def\II{\relax{\rm I\kern-.18em I}}
\def\IK{\relax{\rm I\kern-.18em K}}
\def\IP{\relax{\rm I\kern-.18em P}}
%\def\IX{\relax{\rm X\kern-.01em X}}
%this doesn't work

\font\cmss=cmss10 \font\cmsss=cmss10 at 7pt
\def\IR{\relax{\rm I\kern-.18em R}}

\def\tilde{\widetilde}
\Title{ \vbox{\baselineskip12pt\hbox{hep-th/9706168}
\hbox{RU-97-52}}}
{\vbox{\centerline{The State of Matrix Theory}}}
\centerline{Tom Banks}
\smallskip
\smallskip
\centerline{Department of Physics and Astronomy}
\centerline{Rutgers University }
\centerline{Piscataway, NJ 08855-0849}
\centerline{\tt banks@physics.rutgers.edu}
\bigskip
\bigskip
\noindent
This is a brief description of what has been accomplished and what
remains to be done in the construction of a nonperturbative formulation
of \lq\lq The Theory Formerly Known as String \rq\rq .  It is culled
from two short talks given by the author at SUSY97 and Strings97.

%\draftmode
\Date{June 1997}

\def\mt{Matrix Theory}
\def\nto{N\rightarrow\infty}
\def\cz{{\cal Z}^A}
\def\lp{l_{11}}
\newsec{\bf Introduction}

This is a brief summary of the present state of Matrix Theory given in 
talks at the SUSY 97 conference at the University of Pennsylvania and
the Strings 97 conference in Amsterdam.  The SUSY97
talk was given to an audience dominated by people with no
expertise in String Theory, and technical details were kept to a
minimum.   In addition, since the field is in a rapid state of flux, I
attempted to talk mostly about certain key features of Matrix Theory
which will survive until the time when it is sufficiently well
understood to receive a name which describes what it is really about.
Apart from that I will primarily give only a list of key results. A
fairly extensive reference list can be compiled by downloading citations
of the papers \ref\bfss{T. Banks, W. Fischler, S. Shenker and L. Susskind,
hep-th/9610043, {\it Phys. Rev.}{\bf D55}, (1997), 112.}\ 
\ref\bs{T.Banks, N.Seiberg, hep-th/9702187 } 
\ref\dvv{R.Dijkgraaf, E.Verlinde, H.Verlinde, hep-th/9703030} from the SPIRES
database.

\mt\ is a nonperturbative Hamiltonian formalism for the theory formerly
known as string theory/M theory (more on the ambiguous meaning of the
term M theory below).   Let me begin by describing its most salient
defect.  It is formulated in the infinite momentum frame (IMF)(sometimes
called light cone gauge - though the two concepts are not precisely
equivalent).  Thus, it is not manifestly covariant, nor is it
background independent.  To understand the latter remark, remember the
key feature of IMF kinematics:  the dispersion relation of a particle
like excitation has the form
\eqn\disp{E = \sqrt{P_L^2 + {\bf P}^2 + m^2} \rightarrow \vert P_L \vert
+ {({\bf P}^2 + m^2 )\over 2\vert P_L \vert}}
If we define the light cone Hamiltonian by $H = E - P_L$ then as $P_L
\rightarrow \infty$, we see that particle like excitations with positive
longitudinal momenta have very small light cone energies.  The key
simplification of IMF {\it dynamics} is that one tries to write down a
Hamiltonian which describes only these very low energies (or perhaps
energies which at most 
remain constant as $P_L \rightarrow \infty$).  Degrees of
freedom with negative or vanishing $P_L$ , whose energy blows up in the
limit, are imagined to be integrated out in the Wilsonian manner.

In conventional field theory, a background is defined by the value of a
degree of freedom of the theory which is completely translation
invariant, and in particular, has $P_L = 0$.  
Thus, changes in the background refer to degrees of freedom
which are left out of the conventional IMF description.  It is of course
to be hoped that we will eventually find a covariant description of
Matrix Theory.  However, as I will suggest below that spacetime is
only an approximate concept in \mt\ , it is perhaps premature to define
what we mean by covariant.    

The second key feature of \mt\ is that it is {\it
holographic}\ref\holo{C.B.Thorn, in {\it Proceedings of Sakharov
Conference on Physics}, Moscow, (1991), 447-454, hep-th/9405069; G. 't
Hooft, {\it Dimensional Reduction in Quantum Gravity}, Utrecht preprint
THU-93/26, gr-qc/9310026; L. Susskind, {\it J. Math. Phys.}{\bf 36},
(1995), 6377, hep-th/940989.}.  Imagine
that the longitudinal direction is compactified on a circle, so that
longitudinal momentum is quantized in units ${1\over R}$, with the total
$P_L$ equal to $N/R$ with $\nto$.   A holographic theory contains only
degrees of freedom which carry the smallest unit of longitudinal
momentum (and those connected to them by the gauge transformations to be
described below).   Low energy states corresponding to particles with
finite fractions of the lowest $P_L$ will be composites of these
fundamental objects.  There should thus be at least $N$ of these
fundamental degrees of freedom, which one might think of as the partons
of \mt\ .

It is convenient to arrange these $N$ degrees of freedom as the diagonal
matrix elements of an $N \times N$ matrix ${\cal Z^A}$ ($A$ is a label for
all of the quantum numbers that an individual parton carries.  As we
will see, in compact spacetimes these labels can encode all of the
degrees of freedom of a field theory.).  
If we think of the
matrix elements as the coordinates of partons, then there is a natural
$S_N$ statistics symmetry, which interchanges the partons, and acts on
the matrix $\cz$ as $ \cz \rightarrow S\cz S^{\dagger}$.  
It is tempting to view this as a residuum of a continuous $U(N)$ gauge
invariance, which we can do by adding degrees of freedom to fill out the
diagonal matrix to a general Hermitian matrix.  Thus for example, for a
supersymmetric system of $N$ particles in eleven space time dimensions,
the degrees of freedom will be a nine component vector ${\bf X}$ and a
sixteen component spinor ${\bf \Theta}$, each of whose components is an
$N \times N$ Hermitian matrix.  Conjugation of the matrices by $U(N)$ unitary
transformations is a gauge symmetry, generalizing the $S_N$ of statistics. 
 Remarkably, in the
situations which we understand completely, this simple prescription
(plus IMF Super-Galilean invariance) turns out to completely determine the
interacting theory.  Perhaps this should be viewed as a nonperturbative
generalization of the way in which perturbative string interactions
follow immediately from the laws of string propagation.
It is at any rate a completely different route than that taken in
quantum field theory for generalizing the Fock
space description of free particle propagation to 
an interacting theory.  

The answer to the question, \lq\lq Why don't we see all of these extra,
off diagonal, degrees of freedom in the real world ?\rq\rq , is
remarkable, and comes in several parts.  First we must realize 
that the general setup of holographic theories requires the existence of
bound states of $N$ partons for any value of $N$.  This is simply the
requirement that physical particles come with all possible values of the
longitudinal momentum.  In the examples studied so far, these are always
threshold bound states, so it is not easy to establish their
existence\foot{I am trying to be very general here.  Lack of time and
space precluded the discussion of detailed examples.  The reader should
be assured that they exist and should consult the references mentioned
in the next section. In eleven dimensions the threshold bound state
problem has been solved only for $N=2$ \ref\sethi{S.Sethi, M.Stern, hep-th/9705046.}, 
while for weakly coupled string
limits, it has been solved in general.}.  One way of trying to ensure this is to study
sequences of matrix models which differ only in the dimensionality of
the matrices.  For large $N$, this should guarantee some uniformity of
behavior. 

Let us imagine that we have succeeded in establishing the existence of
some number of such stable bound states in the $N \times N$ matrix model
for every $N$.  Another characteristic feature of the models under
discussion is that nonderivative interactions between the matrices are
always functions of their commutators $[{\cal Z}^A , {\cal Z}^B]$.  
As a consequence,
the classical potential of the model has a large number of flat, zero
energy directions.  Any collection of block diagonal matrices, with
different coefficients of the unit matrix in each block, is a minimum of
the classical potential.  It is extremely important that, as a
consequence of SUSY, these are exact flat directions of the quantum
problem - no effective potential is generated by quantum
corrections\foot{Actually, all that is necessary is that
SUSY is restored at asymptotically large values of the differences
between coordinates in each block.  These differences play the role of
relative separations between particles and asymptotic SUSY is enough to
guarantee that the potential falls off at large distance.}.  
Now let us consider configurations of the system in which all of the
coordinates within each block are put into the wave function of one of
the block bound states.  The coefficient of the unit matrix within this
block acts as the transverse center of mass position of the
corresponding particle.  When the transverse distances between
particles are large, the off (block) diagonal degrees of freedom get
large frequencies (essentially via the Higgs mechanism) 
and are unobservable at low energies (thus answering
the rhetorical question above).  SUSY guarantees that their virtual
effects do not
generate a potential along the flat directions (asymptotic SUSY is
enough to guarantee that any potential falls off with distance).

Thus, we establish the existence of scattering states of arbitrary
numbers of particles.  Put more dramatically, what we have accomplished
here is the systematic derivation of an approximate spacetime picture
from the dynamics of our quantum system.  In the examples which have
been studied, the resulting scattering states are particles and strings
with relativistic dispersion relations - although the underlying
Hamiltonian has only the manifest symmetries of the IMF.  Space is
derived as the moduli space of a supersymmetric quantum system.
Approximate locality (the fact that forces due to virtual effects 
fall off with distance) is a
consequence of SUSY (in bosonic matrix models quantum corrections
generate linear confining potentials along the classical flat
directions).  The apparent singularity of the force law at short
distances is a consequence of improperly integrating out the off
diagonal matrices in a regime where their frequencies are low.

The resulting theory is {\it au fond} nonlocal.  Locality appears as an
approximate property of long distance, low energy interactions.
The original matrix model of \bfss\ describes flat eleven dimensional
spacetime. In a local theory this would immediately have led to a
description of compact spacetimes.  In field theory, degrees of freedom
are associated with individual points.  The difference between a flat
infinite space and a compact one is merely a question of boundary
conditions.  This led to a widespread expectation that the discovery of
eleven dimensional M theory would lead to a complete nonperturbative
formulation of \lq\lq the theory formerly known as string \rq\rq . Thus
there was a conflation of the idea of M theory as just another
asymptotic limit of string moduli space and that of M theory as the
nonperturbative formulation of everything.  

In a nonlocal theory, one may imagine that there are fundamental degrees
of freedom (and not just composite soliton states) associated with
nontrivial topological submanifolds of a compact space.  If, as one
might expect, the energy of such degrees of freedom goes to infinity
with the volume of the submanifold, then the decompactification limit
might not capture all of the degrees of freedom of the theory.
Experience appears to show that this is the case for the Matrix model of
M theory\foot{The cautious phrasing here has to do with the large $N$
limit. For finite $N$ one certainly needs to introduce extra degrees of
freedom to describe compactification.  
I have speculated that these might appear automatically in a
proper organization of the large $N$ limit.  With the advent of clear
evidence that supersymmetric gauge field theories are not the whole
story of compactification, I am preparing to abandon this point of view.
But the evidence is not all in yet.}.  This is closely connected with
the old observation that the duality groups of compactified string
theory grow with the number of compactified dimensions\foot{L.Susskind
has suggested to me that one might have anticipated the existence of
degrees of freedom associated with compact dimensions from the fact that
the entropy of a black hole with a given large mass, increases as the
number of noncompact dimensions is decreased.}. Thus, the
Matrix model has the property that the density of low energy states in
finite volume, {\it increases} as the volume is decreased. I will
discuss
this briefly, along with the rest of the state of the art, in the next
section. 

\newsec{Progress and Problems}

Before beginning the litany of successes of the Matrix model, I want to
make one point.  Many of the successes consist of the reproduction of
facts already known from string duality.  I believe that the correct way
to think about this is by analogy with symmetry arguments in QCD.
String duality should be thought of as the analog of chiral symmetry
while the Matrix Theory (in its final, as yet unrealized form, which
applies to arbitrary compactifications) 
is the analog of the QCD Lagrangian.  It
embodies the symmetries, and fills in dynamical details which cannot be
understood on the basis of symmetry arguments alone.
Time limitations preclude the detailed discussion of even one example.
I will content myself with a list of results and references.

\item{1.} The matrix model describing uncompactified eleven dimensional
spacetime is given by the following supercharges and Hamiltonian
\eqn\supa{Q_{\alpha} = \sqrt{R}\ Tr\ 
\bigl{[} P^i (\gamma^i )_{\alpha\beta} + i [X^i , X^j ]
(\gamma^{ij} )_{\alpha\beta} \bigr{]} \Theta_{\beta} }
\eqn\supb{\tilde{Q}_{\alpha} = {1\over \sqrt{R}}\ Tr\ \Theta_{\alpha} }
\eqn\ham{H = R\ Tr\ \bigl{[} (P^i )^2 - [X^i , X^j ]^2 + \Theta [{\bf \gamma X},
\Theta ] \bigr{]}}
The vectors here are nine dimensional transverse coordinates and their
canonical conjugates, and the
spinors are their sixteen component superpartners.  $R$ is the radius of the
longitudinal direction of the IMF and $N/R$ is the total longitudinal
momentum.  The matrices are $N\times N$.  $\lp$ is the eleven
dimensional Planck length, and has been set to $1$.  
The $N\rightarrow\infty$ limit of 
this model was shown in \bfss\ to contain the Fock space of eleven
dimensional SUGRA, as well as, (following the seminal work of
\ref\dhn{B. de Wit, J. Hoppe, H. Nicolai, {\it Nucl. Phys.}{\bf B305},
[FS 23], (1988),545; P.K.Townsend {\it Phys. Lett.}{\bf B373}, (1996),
68, hep-th/9512062. }) 
large metastable semiclassical membranes.  A low
energy, zero longitudinal momentum graviton scattering amplitude was
calculated and shown to agree with the prediction of SUGRA.  Various
membrane scattering amplitudes \ref\memb{J.Polchinski, P.Pouliot,
hep-th/9704029;  N.Dorey, V.Khoze, M.Mattis, hep-th/9704197.
 O.Aharony, M.Berkooz, {\it Nucl. Phys.}{\bf B491},
(1997), 184, hep-th/9611215; G.Lifschytz, S.Mathur, hep-th/9612087.} 
were calculated including one
with nonzero longitudinal momentum transfer and shown to agree with
SUGRA.  A number of infinite energy BPS brane configurations were found
\ref\branes{T.Banks, N.Seiberg, S.Shenker, {\it Nucl. Phys.}{\bf B490},
(1997), 91, hep-th/9612157; O.J. Ganor, S.Ramgoolam, W.Taylor IV,
hep-th/9611202; M.Berkooz, M.Douglas, {\it Phys. Lett.}{\bf B395},
(1997), 196, hep-th/9610236.}.  
These are best viewed as limits of objects wrapped
around compact directions.  Significantly, although one is able to
exhibit five branes, one only sees those wrapped around the longitudinal
direction\foot{See however the very interesting suggestion for
describing the transverse five brane wrapped on a three torus in Ganor {\it
et. al.} \branes\ .}.  This is related to the failure of the Super Yang Mills
prescription for compactification.

\item{2.}Compactification down to ten dimensions was studied, and in the limit
of small compactification radius, was shown to contain the Fock space of
Type IIA string field theory \ref\motlbs{L.Motl, hep-th/9701025;
T.Banks, N.Seiberg, hep-th/9702187.}.  In particular it is {\it
proven} that the strings become free in the small radius limit. 
This was also done in
\ref\dvv{R.Dijkgraaf, E.Verlinde, H.Verlinde, hep-th/9703030.} 
and there it was shown that the correct leading order string
interactions (including the correct \ref\witten{E.Witten, {\it
Nucl. Phys.}{\bf B443}, (1995), 85, hep-th/9503124.} scaling of the
coupling with the radius) were reproduced by the model.  The procedure
for compactification is to replace the matrices by infinite dimensional
operators. In particular, the compactified coordinates are replaced by
$X^a \rightarrow {\partial \over i\partial\sigma^a } I_{N\times N} -
A_a (\sigma )$, where $A$ is a $U(N)$ gauge potential \bfss\ \ref\taylor{W. Taylor IV,
{\it Phys. Lett.}{\bf B394}, (1997), 283, hep-th/9611042.}\ .  Other variables
simply become matrix valued functions of $\sigma$.  When plugged into
the Hamiltonian \ham\ , this ansatz produces the Hamiltonian of
maximally supersymmetric $1+1$ dimensional super Yang Mills theory
($SYM_{1+1}$).  The zero radius limit forces one onto the moduli space
of this model, which was shown at large $N$ to be the same as the Type
IIA Fock space.

\item{3.} This compactification procedure can be generalized to two and
three dimensional tori, producing $SYM_{d+1}$ for $d = 2,3$.  The theory
has the full duality group of compactified M theory, which arises as a
combination of obvious geometric transformations and electric-magnetic
duality transformations\ref\duality{L.Susskind, hep-th/9611164,
O.J. Ganor, S.Ramgoolam, W.Taylor IV, {\it op. cit.}}.  
In particular, one can study in some detail
the Aspinwall-Schwarz \ref\as{P.Aspinwall, in Proceedings of ICTP
Trieste Conf., Jun. 1995, {\it Nucl. Phys. Proc. Suppl.}{\bf 46},(1996),
30, hep-th/9508154  ; J.Schwarz, {\it Phys. Lett.}{\bf B367}, (1996),
97, hep-th/9510086. } 
limit of compactification on a zero area
two torus, in which a new dimension grows and gives the ten dimensional
Type IIB string theory.  Using the matrix model one can explicitly
derive the rotation symmetry between this new dimension and the existing
noncompact dimensions - something which is completely mysterious from
all other points of view \bs\ \ref\ss{S.Sethi, L.Susskind,
hep-th/9702101. }.  One can also establish that
zero longitudinal momentum ten dimensional graviton scattering emerges
in the correct manner \ref\bfssprime{T.Banks, W.Fischler, N.Seiberg,
L.Susskind, hep-th/9705190.}.  Finally, one obtains a
prediction for wrapped membrane-antimembrane production in graviton
scattering \ref\abss{T.Banks, S.Shenker, {\it In preparation}}.  

\item{3.} In four compact dimensions, the proposed SYM theory is
nonrenormalizable, and also fails to reproduce the correct $SL(5)$
duality group.  The authors of \ref\rbrs{M.Berkooz, M.Rozali, N.Seiberg,
hep-th/9704089, M.Rozali, hep-th/9702136.} argued that a definition of
the $SYM_{4+1}$ theory as a compactification of the $5+1$ dimensional
$(2,0)$ supersymmetric fixed point theory with a spectrum of tensor
charges in the $U(N)$ weight lattice \ref\seib{N.Seiberg,
hep-th/9705117.} repaired all of these
difficulties.  A pattern begins to emerge in which the spectrum of
\lq\lq momenta \rq\rq in the auxiliary quantum theory which defines the 
compactified matrix model, is the spectrum of finite energy BPS charges
coming from branes wrapped around the longitudinal and transverse
directions \ref\bsbrane{T.Banks, N. Seiberg, {\it op. cit.}, T.Banks,
N.Seiberg, S.Shenker, {\it op. cit.}, N.Seiberg, hep-th/9705221.}.  
The new dimension of the $(2,0)$ theory is
associated with the longitudinally wrapped five brane.

\item{4.} Above 4 compact dimensions, arguments have been advanced that
the auxiliary quantum theory can no longer be a local quantum field
theory \ref\brsns{M.Berkooz, M.Rozali, N.Seiberg, {\it op. cit.},
N.Seiberg, hep-th/9705221.}.  It has been identified with the theory of $N$
Type II Neveu-Schwarz fivebranes wrapped around a five dimensional
transverse
torus, in the limit of string theory in which the coupling goes to zero.
Various low energy limits of this as yet rather mysterious theory, as
well as its spectrum of BPS charges, seem to fit the known data.  In
particular, the purely transverse wrapped five brane can be exhibited.
This gives an existence proof, though not yet a construction, of the
matrix model compactified on $T^5$ and perhaps $T^6$.  There is a strong
suspicion that the logarithmic infrared singularities of a holographic
theory with only two noncompact transverse dimensions will drastically
change
the story of compactification on spaces of dimension $7$ and higher.
Thus, compactifications with four or fewer large spacetime dimensions
are beyond the limits of current knowledge.

\item{5.} All of the above refers to compactifications with maximal
SUSY. Compactification of M theory with half maximal SUSY down to ten
dimensions has been successfully carried out
\ref\dfetal{U.H. Danielsson, G.Ferretti, hep-th/9610082; S.Kachru,
E.Silverstein, {\it Phys. Lett.}\break\ {\bf B396}, (1997), 70, hep-th/9612162;
L.Motl, hep-th/9612198; S.J.Rey, T.Banks, N.Sei-\break berg, E.Silverstein,
hep-th/9703052; T.Banks, L.Motl, hep-th/9703218; D.A.Lowe, 
hep-th/9702006, hep-th/9704041; S.J.Rey, hep-th/9704158; P.Horava, hep-th/9705055. }.  Heterotic
string theory has been derived in the ten dimensional limit, and the
Horava-Witten \ref\hw{P.Horava, E.Witten, {\it Nucl. Phys.}{\bf B475}
(1996), 94, hep-th/9603142.} picture of its strong coupling limit has been
verified by explicit matrix model calculations.  Compactification on
further circles in this formalism leads to unresolved problems
\ref\bmh{T.Banks, L.Motl {\it op. cit.}, P.Horava, {\it op. cit.}} 
with anomalies.  There are strong indications that these
problems can be resolved, and a full picture of the moduli space
obtained, only by using the $(2,0)$ fixed point theory compactified on
$K3\times S^1$ and taking various limits \ref\br{S.Govindarajan,
hep-th/9705113; M.Berkooz, M.Rozali, hep-th/9705175.}. 
There exists a proposal for compactifications on five dimensional
manifolds
with half maximal SUSY via the weak coupling limit of heterotic
Neveu-Schwarz fivebranes \ref\ns{N.Seiberg, hep-th/9705221} (see also
\ref\tfivemodztwo{A. Fayyazudin, D.J.Smith, hep-th/9703208; N.Kim,
S.J. Rey, hep-th/9705132.}).  

\item{6.} Finally, I would like to note that the work of
\ref\dos{M.Douglas, H.Ooguri, S.Shenker, hep-th/9702203;
 W.Fischler, A.Rajaraman, hep-th/9704086. }\ shows that when there are no
strong SUSY nonrenormalization theorems, correct answers in Matrix
Theory will only be obtained after the large $N$ limit is taken.  This
points up the need for some sort of large $N$ renormalization group to
isolate the dynamics of those states with energies of order $1/N$.  This
is very different from standard large $N$ physics (essentially because
of the SUSY flat directions).  At the moment, we only have a general
treatment of this problem in the weakly coupled string theory limit,
where it is solved by the renormalization group of two dimensional field
theory .

\newsec{\bf Conclusions}

It is fairly clear that Matrix theory is a nonperturbative formulation
of a theory which underlies string theory.  It is also fairly clear that
this theory is completely nonlocal in nature, and incompletely
understood. Spacetime is only an approximate concept in the theory and
it is clear that much of our difficulty in formulating it is related to
our reliance on spacetime intuitions.  More fundamental to the theory is
the supertranslation algebra, including charges associated with wrapped
two branes and fivebranes.  Some of the approximate space-time limits of
the theory utilize these brane charges as momenta, so it is clear that
we should treat them on an equal footing with ordinary momenta.  Perhaps
the ultimate shape of the theory will be related to allowed
representations of this algebra.  The charge spectrum is generally
constrained to lie in a lattice, via arguments related to the Dirac
quantization condition.  We should find a derivation of these
constraints which does not rely on spacetime notions.  I would suggest
that ultimately, compactifications will be described simply in terms of
the allowed spectra of BPS charges \foot{The first hint that other BPS
charges should be considered on an equal footing with momentum comes
from perturbative T duality.  The work of Aspinwall and Schwarz cited
above gave another indication of this, as did various other results in
string duality.  P.K. Townsend \ref\pkt{P.K.Townsend, hepth/9507048.} 
was the first to advocate the
importance of the full M theory SUSY algebra, and I. Bars
\ref\bars{I. Bars, hepth/9604139.}
was the first to emphasize that all charges should be put on the same
footing as momenta. Hints of this approach to geometry can be found in
the last section of \bfss\ }.    P. Aspinwall
and D. Morrison have assured me that at least for complex dimensions two and
three, the energies of BPS states associated with supersymmetric cycles 
determine the geometry of a Calabi Yau
manifold up to a few finite choices.
I am suggesting that in the final formulation of the theory, 
there will be no other measure of geometry which
is unconnected with some sort of low energy approximation.

Matrix theory is undergoing very rapid development.  I believe strongly
that our view of it a year from now will be radically different from
anything we are now saying.  I believe that the supertranslation
algebra, the general construction of multibody states from block
diagonal matrices, and the concept of space as the moduli space of a
SUSY quantum system will remain, but that much of the rest of what has
been done up till now will be seen as a series of special
approximations.  I am virtually certain that we will have to understand
the large $N$ limit much more throughly than we do today.
I suspect that the field theoretically unheard of way
in which the density of low energy states changes with the volume will
lie at the heart of the black hole information paradox and the problem
of the cosmological constant.  I am curious to see what the covariant
formulation of the Matrix Theory will look like, and how cosmology gets
into the picture (more than ever this seems to be a question of the
dynamics of moduli).  I wonder whether the apparent existence of a global
definition of time is an artifact of the light cone gauge or a sign that
our notions of general covariance are wrong (almost inevitable if
spacetime does not really exist).  And does the fact that we live in
four large dimensions have to do with the fact that in a holographic
theory, two noncompact space dimensions is the first place where
transverse potentials grow at infinity?  

My purpose in this talk has been to show you that something exciting is
going on.  As string theorists we are used to this happening every
couple of years.  At the risk of seeming immodest, I would like to
propose that the present situation has within it the seeds of something 
more striking than anything we've seen before.  In our study
of various limits of moduli space we've succeeded in coding much of what
we learn into low energy effective field theories.  This beautiful and
efficient tool has also been the means which allowed us to squeeze
string theory into our low energy prejudices about geometry and
locality.  Matrix theory, as ugly and unwieldy as it is, has the virtue
of refusing to be forced into this mold.  Instead it forces us 
to give up our blinders and face the bizarre nature of the beast we have
snared straight on.  To the present author at least it seems quite clear
that the fundamental rules of the new theory will seem outlandish to
anyone with a background in quantum field theory or general relativity.
The challenge will be to find the correspondence principle by which the
old rules of geometry and locality emerge from these new axioms.  At the
moment it appears that the only things which may remain unscathed are
the
fundamental principles of quantum mechanics.  
I invite you to join in the search for these rules.  It's lots of fun,
and it leaves your mind panting like a long distance runner after a
marathon. Maybe in a few years we'll even find a decent name for the
damned thing.

\centerline{\bf Acknowledgements}
I would like to thank all of my collaborators, W.Fischler, S.Shenker,
L.Susskind, N.Seiberg, L.Motl, E.Silverstein, and a host of other
authors for sharing the adventure of Matrix Theory with me.   This work
was supported in part by DOE grant \# DE-FG02-96ER40559.
\listrefs
\end